    \documentstyle[epsf]{article} \catcode`\@=11
    \def\section{\@startsection{section}{1}{\z@}%
    {-3.5ex plus -1ex minus -.5ex}{1.5ex plus.3ex}{\bf }}
    \def\subsection{\@startsection{subsection}{1}{\z@}%
    {-3.5ex plus-1ex minus-.5ex}{1.5ex plus.3ex}{\bf }} 
    
\begin{document}
    \vspace*{-1.14cm}\hfill\parbox{5.40cm}{\large Leipzig Preprint 
                                        NTZ 29/1998}\\[0.3cm]\mbox{}
    \hfill\parbox{4.77cm}{\Large\centering Annalen\\der
    Physik\\[-.2\baselineskip] {\small \underline{\copyright\ Johann
    Ambrosius Barth 1998}}} \vspace{.75cm}\newline{\Large\bf
Energy barriers of spin glasses from multi-overlap simulations
    }\vspace{.4cm}\newline{\bf   
Wolfhard Janke$^{1,2}$, Bernd A. Berg$^{3,4}$ and Alain Billoire$^5$
    }\vspace{.4cm}\newline\small
$^1$Institut f\"ur Theoretische Physik, Universit\"at Leipzig, 
D-04109 Leipzig, Germany\\
$^2$Institut f\"ur Physik, Johannes Gutenberg-Universit\"at, D-55099 Mainz,
Germany\\
$^3$Department of Physics, The Florida State University,
                      Tallahassee, FL~32306, USA\\  
$^4$Supercomputer Computations Research Institute,
                      Tallahassee, FL~32306, USA\\
$^5$Service de Physique Th\'eorique de Saclay, F-91191 Gif-sur-Yvette, France\\
{\tt wolfhard.janke@itp.uni-leipzig.de, berg@hep.fsu.edu, 
billoir@spht.saclay.cea.fr}
    \vspace{.2cm}\newline 
Received 12 October 1998, revised version 12 October 1998, accepted 12
October 1998
    \vspace{.4cm}\newline\begin{minipage}[h]{\textwidth}\baselineskip=10pt
    {\bf  Abstract.}
We report large-scale simulations of the 
three-dimensional Edwards-Anderson Ising spin glass
system using the recently introduced multi-overlap
Monte Carlo algorithm. In this approach the temperature is fixed and
two replica are coupled through a weight factor such that a broad 
distribution of the Parisi overlap parameter $q$ is achieved. Canonical 
expectation values for the entire $q$-range (multi-overlap) follow by 
reweighting. We present an analysis of the performance of the algorithm 
and in particular discuss results on spin glass free-energy barriers 
which are hard to obtain with conventional algorithms. In addition we
discuss the non-trivial scaling behavior of the canonical $q$-distributions 
in the broken phase.
    \end{minipage}\vspace{.4cm} \newline {\bf  Keywords:}
Ising spin glass; Multi-overlap simulations; Energy barriers; Finite-size
scaling
    \newline\vspace{.2cm} \normalsize
\section{Introduction}
The intuitive picture for spin glasses and other systems with conflicting
constraints, for reviews see~\cite{reviews}, is that there
exists a large number of degenerate thermodynamic states with the same
macroscopic properties but with different microscopic configurations. These
states are separated by free-energy barriers in phase space, caused by
disorder and  frustration. Experimentally this is supported by long
characteristic times found in phenomena like the measurement of the
remanent magnetization in typical spin glasses 
such as, e.g., 
$({\rm Fe}_{0.15} {\rm Ni}_{0.85})_{75} {\rm P}_{16} {\rm B}_6 {\rm Al}_3$
\cite{Gr87}.
However, one difficulty of the theory of spin
glasses is to give a precise meaning to this classification:
No explicit
order parameter exists which allows to exhibit the barriers. The way out
of this problem 
appears to use an implicit parametrization, the Parisi
overlap parameter $q$, which allows to visualize at least some of 
them~\cite{reviews}.
Calculations of the thus encountered barriers in $q$ are of major interest.
For instance, it is unclear whether the degenerate thermodynamic states
are separated by infinite barriers or whether this is just an artifact of
mean-field theory.

Before performing numerical calculations of these barriers, one of
the questions which ought to be addressed is ``What are suitable weight
factors for the problem?'' The weight factor of canonical Monte Carlo (MC)
simulations is $\exp (-\beta E)$, where $E$ is the energy of the
configuration to be updated and $\beta$ is the inverse temperature in
natural units. The Metropolis and other 
MC methods generate canonical
configurations through a Markov process. However, by their very definition
the rare-event states associated with the
free-energy barriers are suppressed in such an ensemble. Now, it
became widely recognized in recent years that MC simulations with a-priori
unknown weight factors, like for instance the inverse spectral density
$1/n(E)$, are also feasible and deserve to be considered, for reviews see
\cite{muca}. Along such lines progress has been made by
exploring~[4--7] innovative weighting
methods for the spin glass problem.

The main idea of the
studies~[4--7] is to avoid getting stuck
in metastable low-energy states by using a Markov process which samples
the ordered as well as the disordered regions of configuration space in
one run. Refreshing the system in the disordered phase clearly benefits
the simulations, but the performance has remained below early expectation.
One reason appears to be that the direct ({\it i.e.} ignoring the
dynamics of the system)  barrier weights are not affected, such that
the simulation slows down due to the tree-like structure of the low-energy
spin glass states, see Ref.~\cite{BhSe97} for a detailed discussion.

In Ref.~\cite{bj98} two of the authors introduced a method which focuses
directly on enhancing the probability for sampling the barrier regions
in the Parisi overlap parameter. Relying
on this method, we have 
meanwhile performed large-scale simulations of
the three-dimensional Edwards-Anderson Ising spin glass. First results
from this novel investigation are reported in the following. The next
section briefly introduces the spin glass model and is followed by
an introduction to the method of~\cite{bj98} in section~3. Technical
details of the computer simulation are described in section~4 and some
of the obtained physical results are summarized in section~5, followed
by conclusions in the final section~6.
\section{Model}
We focus on the three-dimensional ($3d$) Edwards-Anderson Ising (EAI) 
spin glass on a simple
cubic lattice of size $N = L^3$ with periodic boundary conditions. It is 
widely considered to be the simplest model to exhibit realistic spin glass 
behavior and has been the testing ground of 
Refs.~[4--7]. The energy is given by
\begin{equation} \label{energy}
E = - \sum_{\langle ik \rangle} J_{ik}\, s_i s_k \qquad ,
\end{equation}
where the sum is over nearest-neighbour sites and the Ising spins $s_i$
and $s_k$ take values~$\pm 1$. The exchange coupling constants $J_{ik}$ 
are quenched random variables which are chosen to be $\pm 1$ with equal 
probability.
Each fixed assignment of the exchange coupling constants $J_{ik}$ defines a 
realization of the system, and all physical results refer to an average 
over many such realizations. 
Early MC simulations of the $3d$ EAI model
located the freezing temperature at
$\beta_c \approx 0.9$. For a concise review, see Ref.~\cite{MaPaRu97}.
Recent, very high-statistics canonical
simulations~\cite{KaYo96} estimate $\beta_c=0.901\pm~0.034$, and were
interpreted~\cite{MaPaRu97} to improve the evidence in support of a 
second-order phase transition at $\beta_c$.

The Parisi overlap parameter is defined as~\cite{reviews}
\begin{equation} \label{q}
q = {1\over N} \sum_{i=1}^N s_i^1 s_i^2\qquad ,
\end{equation}
where the spins $s^1_i=\pm 1$ and $s^2_i=\pm 1$ correspond
to two independent copies (replica) of the same realization (defined by its 
couplings $J_{ik}$), each with its own time evolution in the MC simulation
(realized by different random numbers). 
\section{Multi-overlap algorithm}
The method of Ref.~\cite{bj98} simulates two replica of the same
realization in one computer run. In another context this has before
been done in~\cite{KeRe94}. Our basic observation, closely
related to multicanonical methods~\cite{muca,BeCe92}, is that one 
does still control canonical expectation values at temperature 
$\beta^{-1}$ when one simulates with a weight function
\begin{equation} \label{weight}
 w(q) = \exp \left[ \beta \sum_{\langle ik \rangle} J_{ik}
(s^1_i s^1_k + s^2_i s^2_k) + S(q) \right] \qquad .
\end{equation}
Of
particular interest is to determine $S(q)$ recursively
such that the 
$q$-distribution
becomes uniform in $q$ (``multi-overlap''), 
and the interpretation of $S(q)$ being the microcanonical entropy of the
Parisi order parameter. Hence, although an explicit order parameter does
not exist, an approach very similar to the multimagnetical~\cite{BeHaNe93}
(which is an highly efficient way to sample interface barriers for
ferromagnets) exists herewith.

As a measure of the performance of the algorithm we recorded the average 
number of sweeps that are necessary to perform tunneling events of the
form 
$$ (q=0) \to (q=\pm 1) ~~{\rm and\ back}.$$
In the following we refer to this quantity as autocorrelation time.
\section{Simulation}
\begin{figure}[t]
\vspace{10.2cm}
\includegraphics{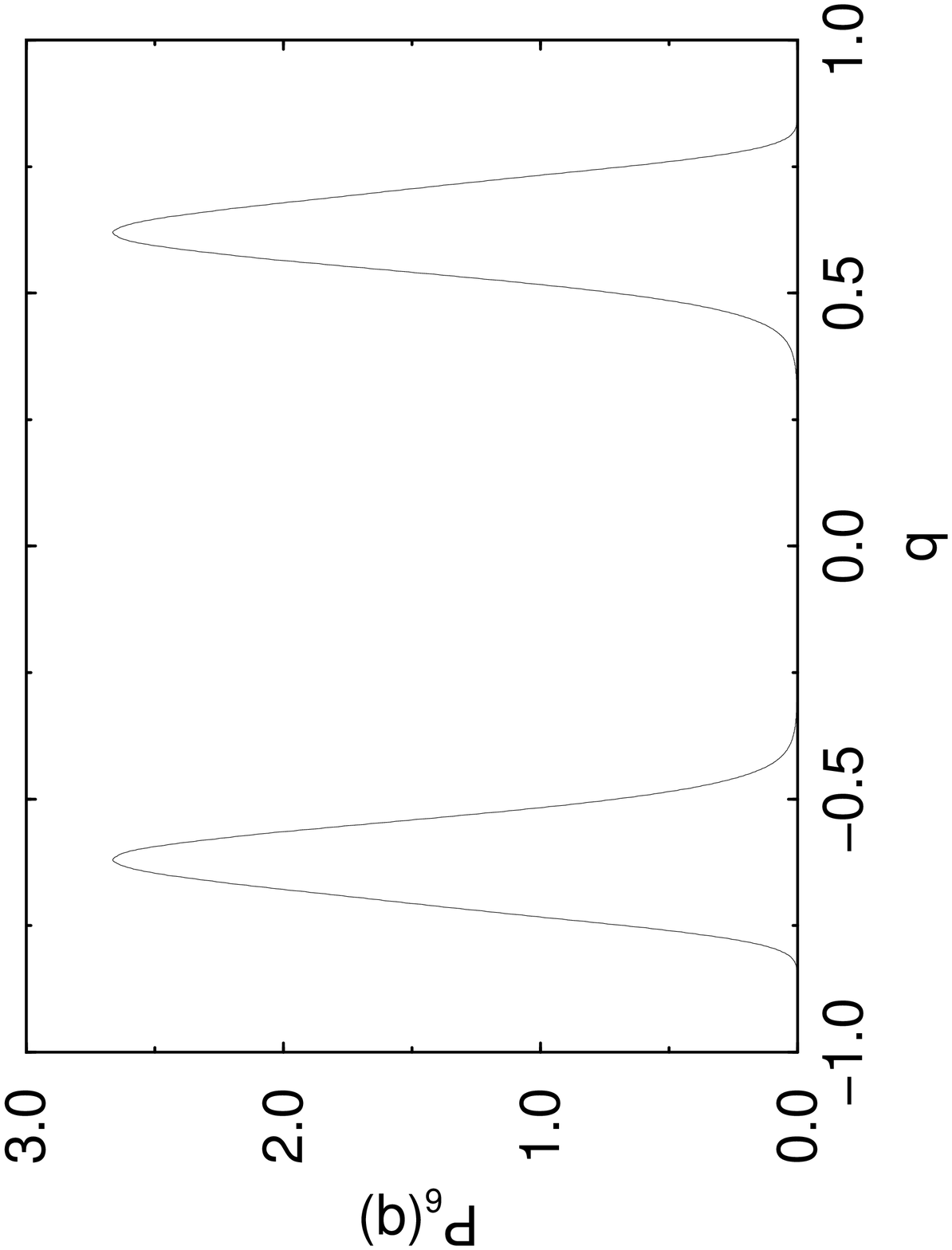}
\includegraphics{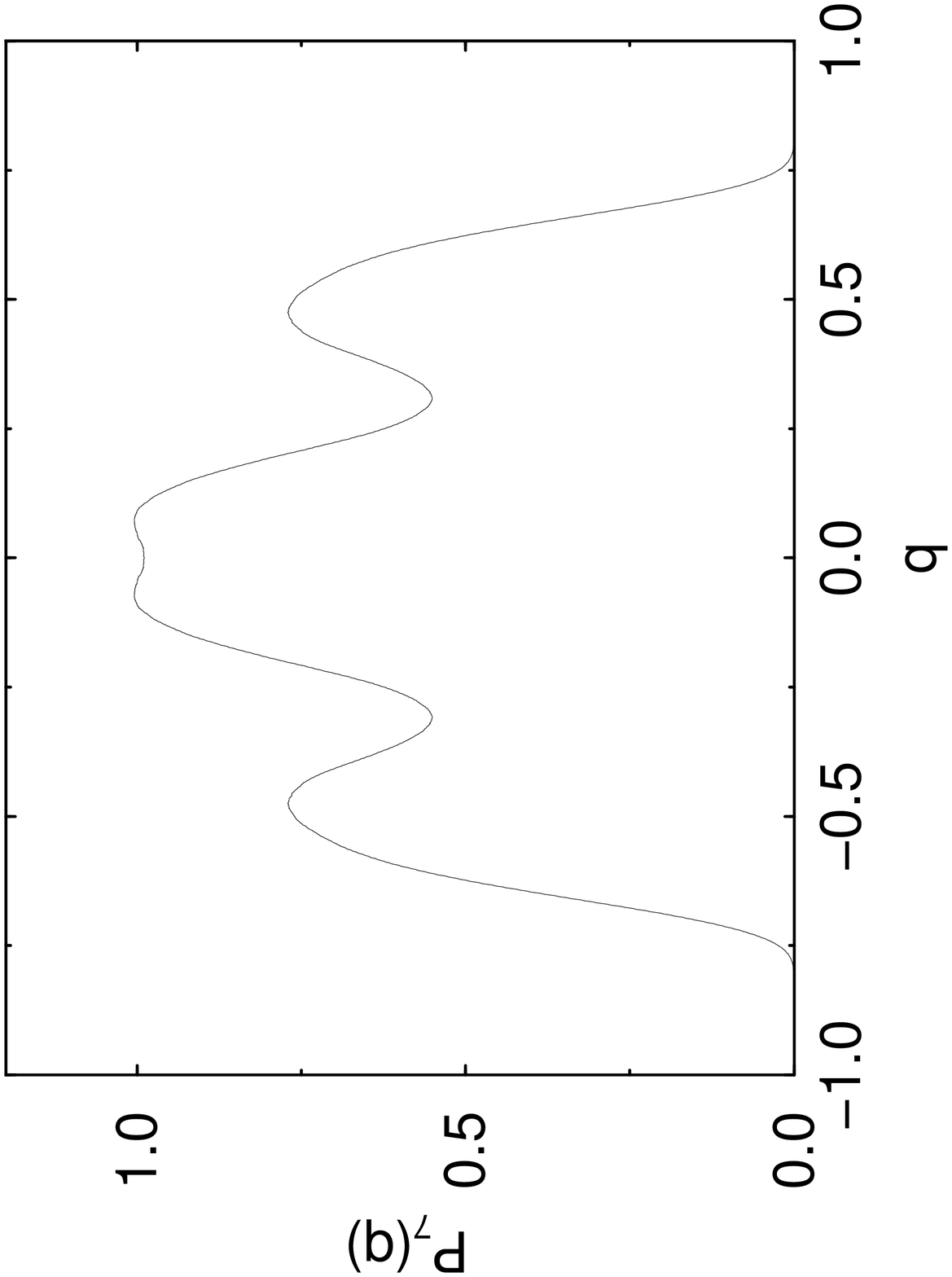}
\includegraphics{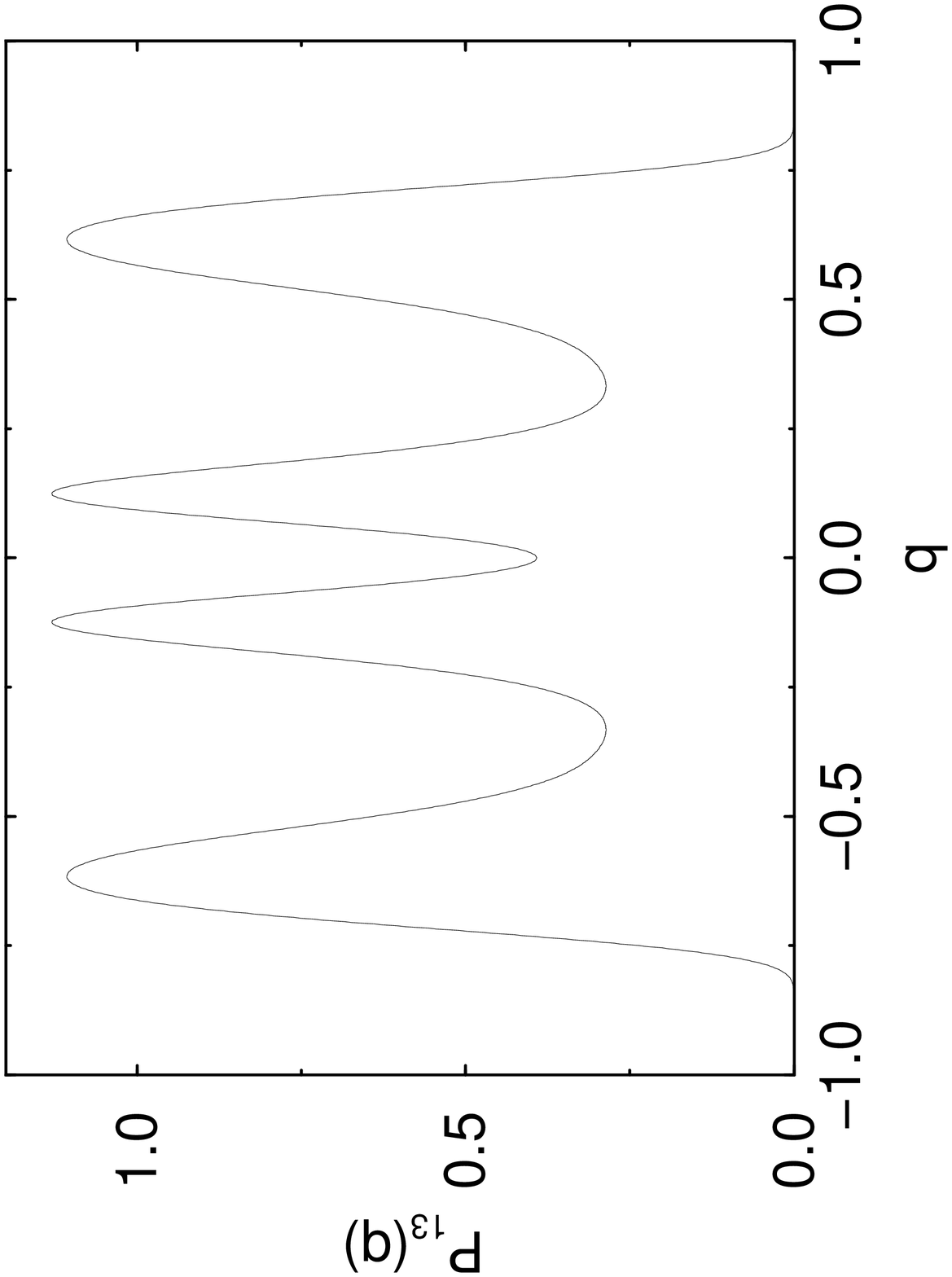}
\includegraphics{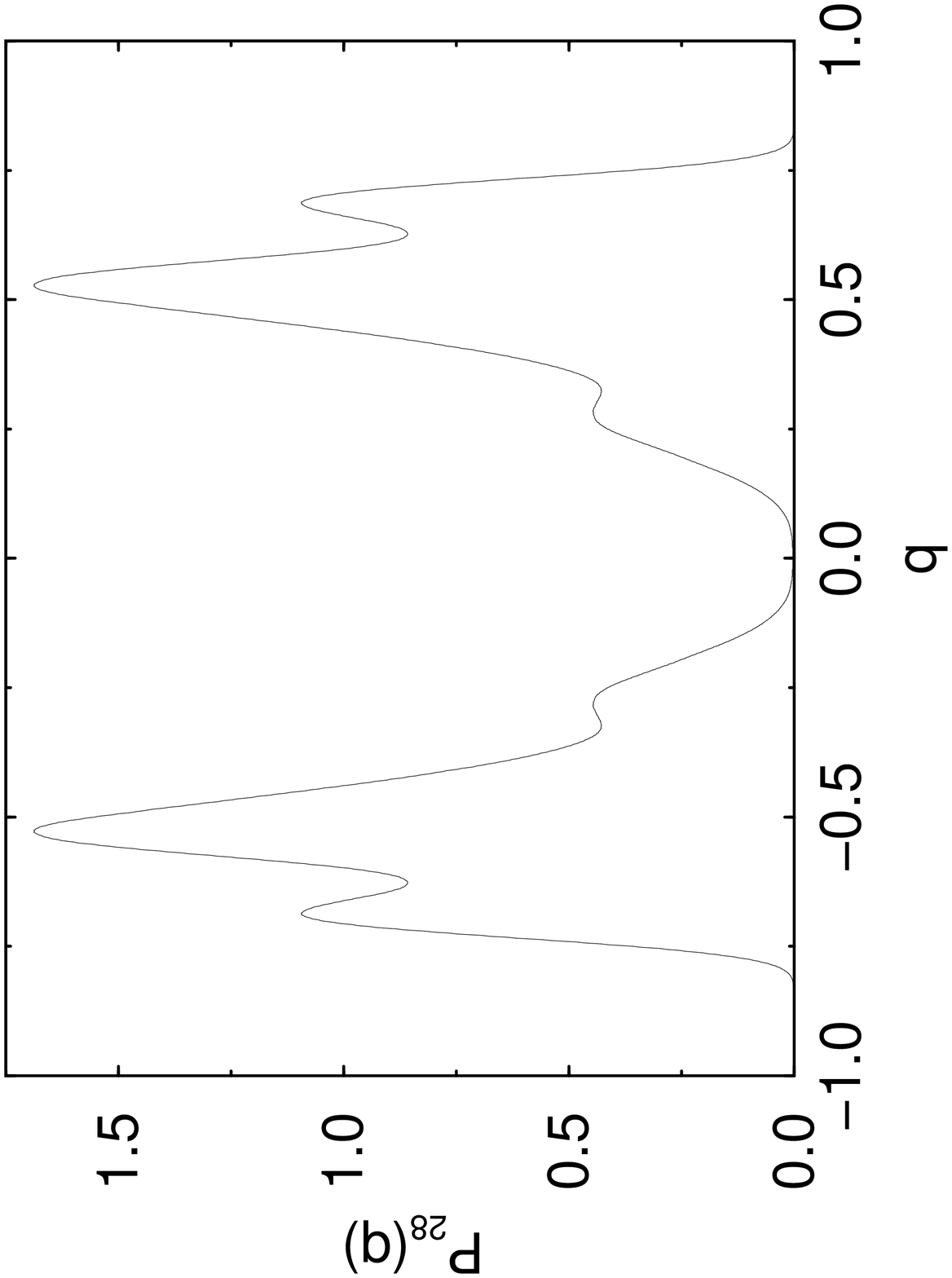}
  \caption{\protect\small  Zoo of probability densities $P_i(q)$ for
                           the $12^3$ system at $\beta=1$. The
                           subscript $i=1,\dots,512$ labels the
                           different realizations.
    }
  \label{fig:PJs}
\end{figure}

The multi-overlap algorithm is thus particularly designed for simulations
of the interesting region 
below the freezing temperature where
$P_i(q)$, 
the canonical probability density of $q$
for realization $i$ (additional dependence on lattice size and 
temperature is implicit),
exhibits a rather complicated behavior. The shape of $P_i(q)$ 
ranges from a simple 
double-peak structure to involved structures of several minima and maxima. 
A few examples encountered in the simulations of the $12^3$ system 
are shown in Fig.~\ref{fig:PJs}.
This is the situation which is notoriously difficult to study with standard 
algorithms. In our EAI study we therefore focussed on simulations at
$\beta = 1 > \beta_c \approx 0.9$. We investigated lattices of size $N = L^3$ 
with $L = 4$, 6, 8, and 12, using T3E parallel computers in Grenoble, Berlin, 
and J\"ulich. We simulated 4096 different realizations for the smaller 
systems up to $L=8$, and 512 for
$L=12$, using the Marsaglia random number generator RANMAR~\cite{Ma90} 
for drawing the $J_{ik}$. As a check we also simulated a second set of 
4096 realizations for $L=4$, 6, and 8, where the $J_{ik}$ were generated 
with 
a more elaborate version of RANMAR: RANLUX~\cite{Lu94} with luxury 
level 4. In the simulations themselves we always employed the RANMAR 
generator due to CPU time considerations.

For each realization the simulation consisted of three steps:
\begin{enumerate}
\item Construction of the weight function (\ref{weight}). Here we employed 
an improved variant of the accumulative iteration scheme discussed 
in Ref.~\cite{Be96}. Details will be published elsewhere. The iteration 
was stopped after at least 10 tunnelings (between $4-20$ for $L=12$)
occurred. 
\item Equilibration run. This run was to equilibrate the system for
given fixed weight factors.
\item Production run. Each production run of data taking was concluded 
after at least 
20 tunneling events were recorded. To allow for
easy (standard) reweighting in temperature we stored besides 
the Parisi overlap parameter also the energies and magnetizations of
the two replica in a time-series file. Since the corresponding 
autocorrelation times $\tau$ vary significantly from realization to 
realization we used an adaptive data compression routine to make sure 
that the spacing between measurements was always proportional to $\tau$.
Using this adaptive scheme we recorded for each realization 65536
measurements.
\end{enumerate}

Let us conclude this section with a few remarks on the performance of the
new algorithm.
Fitting the estimates of the mean autocorrelation time $[\tau]_{\rm av}$
to the form
$ \ln ([\tau]_{\rm av}) = a + z\, \ln (N) $ gives
$z=2.42\pm 0.03$.
As will be discussed below, the implied improvement with respect to
barrier calculations is huge. Nevertheless the slowing down is quite
off from the theoretical optimum, which is $z=1$ for multicanonical
simulations \cite{muca}. 
One reason seems to be that we want the 
$q$-distribution
to be uniform in the whole admissible interval $q \in [-1,+1]$, including the
region of $|q| \approx 1$ that is strongly correlated with ground states and 
hence difficult to reach by local updates, see for instance \cite{BhSe97}.
Being content with a smaller region (like the region between the
two outmost maxima of $P_i(q)$) is
expected to give further improvements of the tunneling performance.
By also monitoring the encountered minimum, maximum and
median tunneling times we observed that the mean values are
systematically larger than the median, what means that the tunneling
distribution has a rather long tail towards large tunneling times.
On the other hand, the effect is not severely hindering our
multi-overlap simulations: For the lattice sizes $L=4$ to 8 the
worst behaved realization took never more than 1\% of the entire
computer time. Indicating a remarkable increase of complexity,
this amount was about 12\% for $L=12$.
\begin{figure}[t]
 \vspace{6cm}
\includegraphics{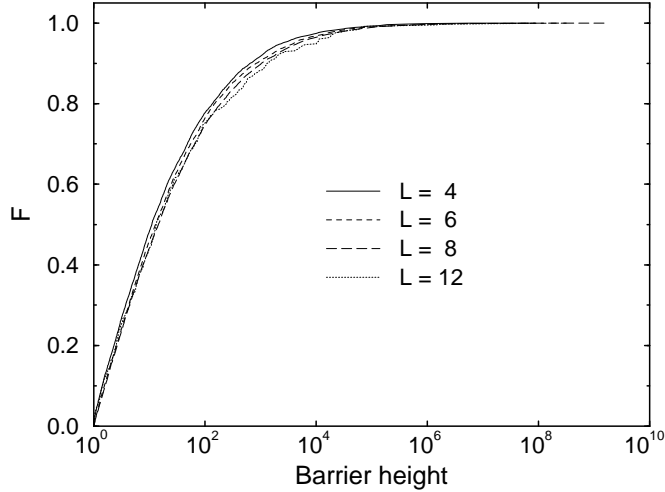}
  \caption{\protect\small 
                          Integrated probability density $F$ of canonical 
                          tunneling barrier heights at $\beta = 1$.
    }
  \label{fig:barrier}
\end{figure}
\section{Results}
The data created in this way allows us to calculate a number of
physically interesting quantities. Let us first concentrate on
the canonical potential barriers in $q$ which were in Ref.~\cite{bj98}
defined as 
\begin{equation} \label{barrier}
B_i = \prod_{q=-1}^{-\, \triangle q} \max \left[ 1,
      P_i(q)/P_i(q+\triangle q) \right]\qquad ,
\end{equation}
where $\triangle q$ is the step-size in $q$. 
For the double-peak
situations of first-order phase transitions \cite{BeHaNe93}
Eq.~(\ref{barrier}) simplifies to $B_i=P_i^{\max}/P_i^{\min}$,
where $P_i^{\max}$ is the absolute maximum and $P_i^{\min}$ is the
absolute minimum (for ferromagnets at $q=0$) of the probability
density $P_i(q)$. Our definition generalizes to the situation where
several minima and maxima occur due to disorder and frustration.
When evaluating (\ref{barrier}) from numerical data for $P_i (q)$
some care is needed to avoid contributions from statistical
fluctuations of $P_i(q)$.

Graphically, our values for the $B_i$ are presented in 
Fig.~\ref{fig:barrier}. It comes as a surprise that the finite-size 
dependence of the distributions is very weak.
To study this issue further, we have
compiled in Table~\ref{tab:bar} for each lattice size the following 
informations
about our potential barrier results: largest and second largest
values $B_{\rm max}$ and $B_2$, median values $B_{\rm med}$ and mean 
values $[B]_{\rm av}$ with their statistical error bars. From this table 
it becomes obvious, why this investigation could not be performed using
canonical methods to which in this context also multicanonical
simulations and enlarged ensembles belong, as their weights for
those barriers are still canonical. For these methods
the slowing down is proportional to the average barrier height
$[B]_{\rm av}$, which is already large for $L=4$, about 17 thousand, and
increases to about 250 thousand for $L=8$. The subsequent decrease to
about 150 thousand for $L=12$ has to be attributed to the smaller number
of realizations in this case: Comparison of the extrem values makes
only sense when the numbers of realizations match. On $L=4$ and $L=6$ 
systems
we have performed a number of (very long) canonical simulations to estimate
the proportionality constant between barrier height and improvement due
to our multi-$q$ simulations. Using these result, we estimate that with
our computer program a canonical MC calculation of the worst $L=12$ barrier
alone, the one reported in Table~\ref{tab:bar}, would take about 1000 years 
on a 500~MHz Alpha processor.
%
\begin{table}[t]
\caption{
\label{tab:bar}
Canonical potential barriers: maximum (and its contribution
to the mean in \%), second largest value, the median and its
jackknife error bar, the mean and its (standard) error bar.}
\vspace{0.3cm}
\centering
\begin{tabular}{ccccc}                               
\hline\hline
$L$& $B_{\max}$     & $B_2$ & $B_{\rm med}$&$[B]_{\rm av}$\\
\hline
 4 & 8.76E07 (62\%) &6.01E06 & $11.05\pm 0.24$ & $(1.72\pm 1.08)$E04\\
 6 & 3.73E08 (41\%) &3.63E08 & $12.83\pm 0.45$ & $(1.12\pm 0.64)$E05\\
 8 & 1.54E09 (77\%) &1.41E08 & $14.52\pm 0.26$ & $(2.44\pm 1.89)$E05\\
12 & 9.14E07 (73\%) &1.78E06 & $13.29\pm 0.88$ & $(1.53\pm 1.44)$E05\\
\hline\hline
\end{tabular}
\end{table}

The reader may be puzzled by the very large error bars assigned
to the mean barrier values. Their explanation is: The entire
mean value is dominated by the largest barrier, which contributes
between 41\% $(L=6)$ and 77\% $(L=8)$, see $B_{\max}$
in the second column of Table~\ref{tab:bar}. Besides $B_{\max}$, 
the second largest value $B_2$ is listed in the third column.
It may be remarked that most of these worst
case barriers exhibit simple double-peak behavior. One lesson from 
these numbers is that very few of the realizations are
responsible for the collapse of canonical simulation methods.

Typical realizations, described by the median results of Table~\ref{tab:bar},
have much smaller tunneling barriers. They turn out to be quite
insensitive to the lattice size.
This result of an almost constant typical
tunneling barrier is consistent with the fact
that our tunneling times are rather far apart from their
theoretical optimum: Still other reasons than overlap barriers have to
be responsible. Therefore, one may question the apparently accepted
opinion that these are the typical barriers which are primarily 
responsible for the severe slowing down of canonical MC simulations.

\begin{figure}
\vspace*{6.8cm}
\includegraphics{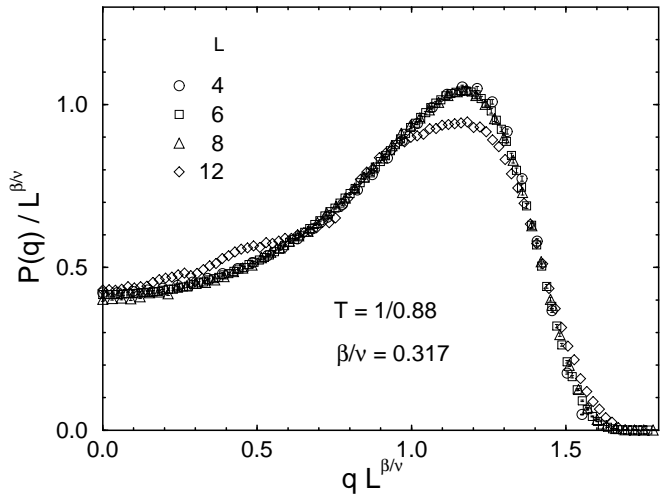}
  \caption{\protect\small
Finite-size scaling plot for $P(q)$ at the transition temperature.
}
\label{fig:P088scal317}
\end{figure}

\begin{figure}[th]
\vspace*{7.0cm}
\includegraphics{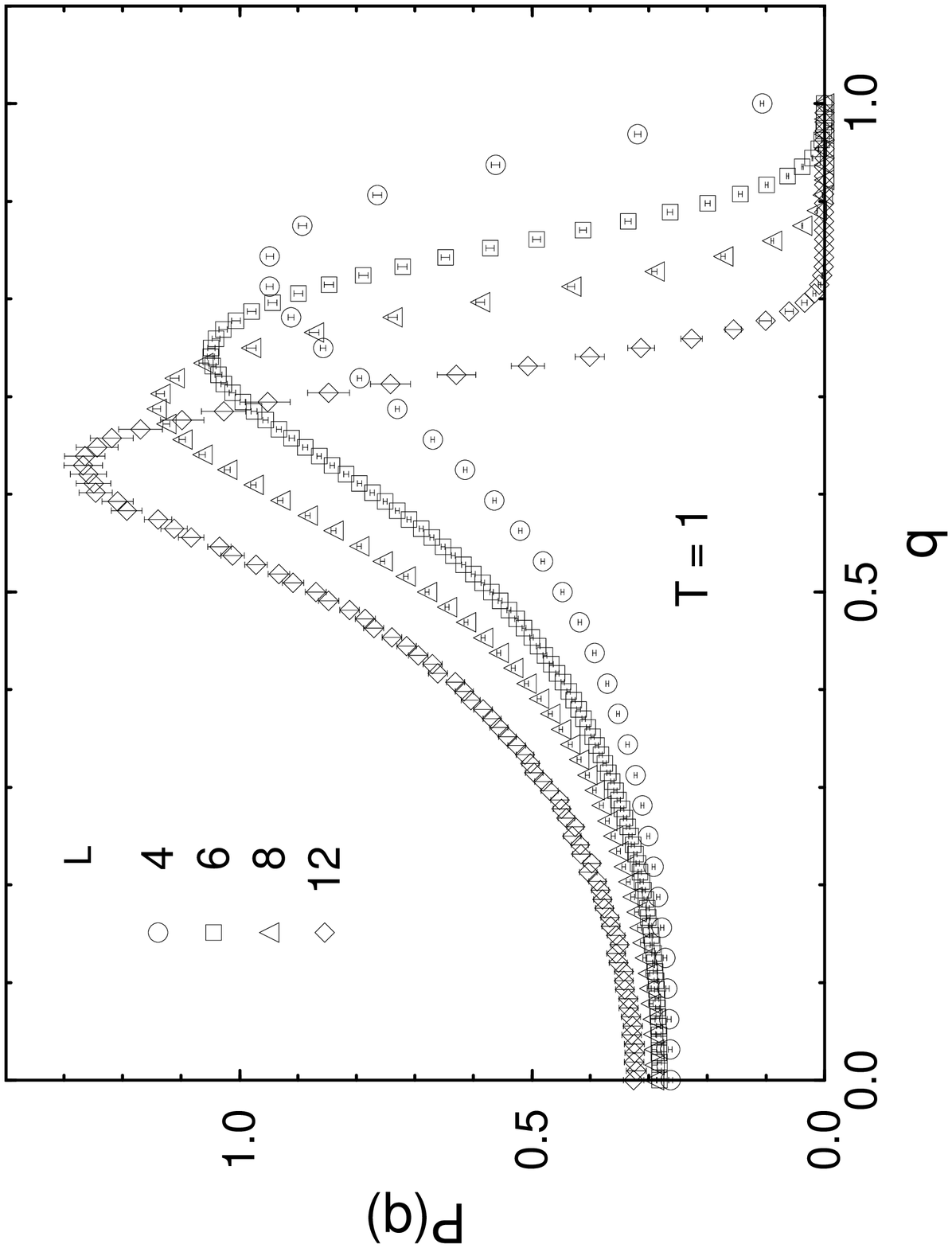}
  \caption{\protect\small
Probability densities $P(q)$ in the spin-glass
phase at $T \approx 0.88 T_c$.
}
\label{fig:P100}
%
\vspace*{7.5cm}
\includegraphics{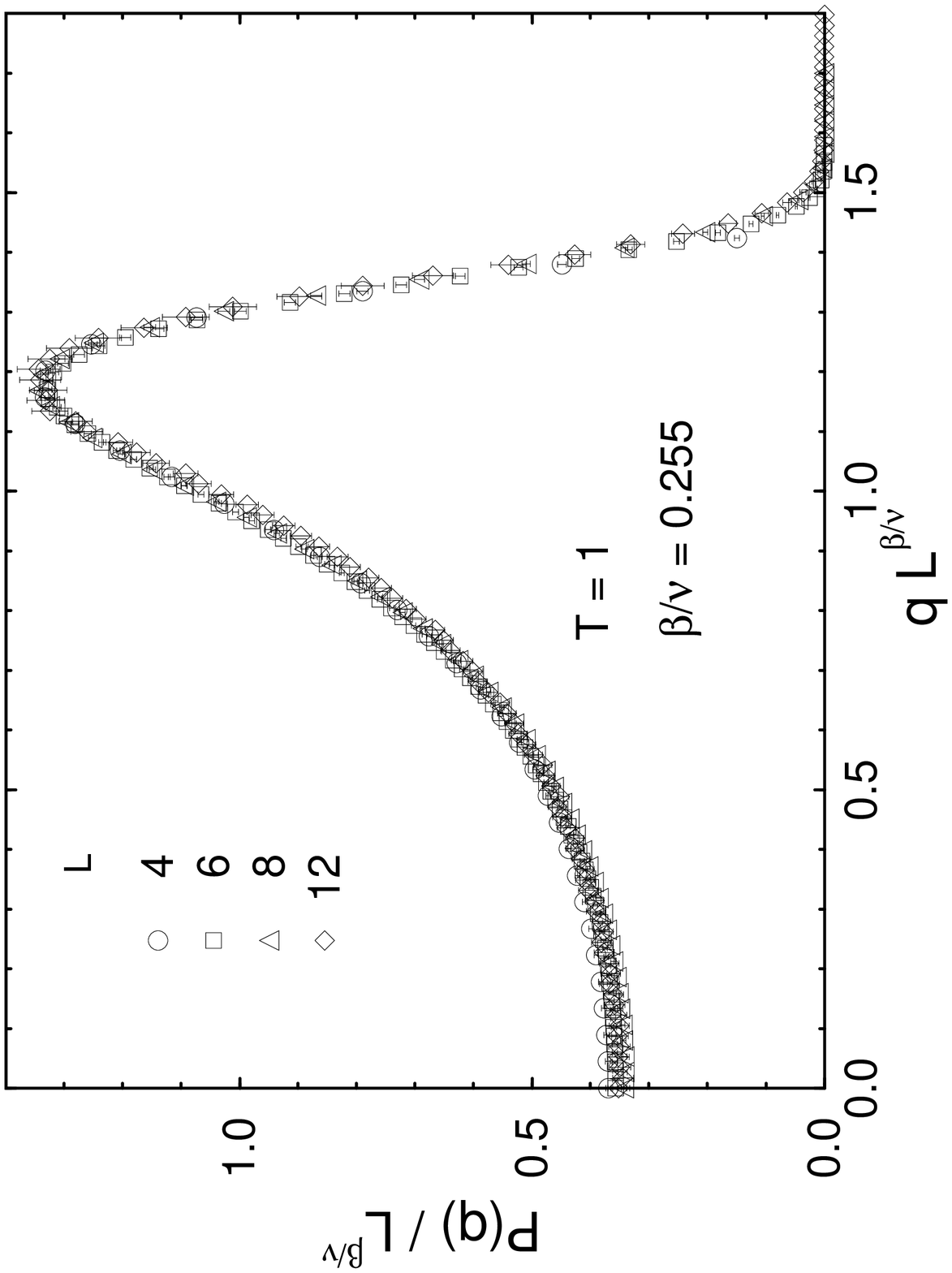}
  \caption{\protect\small
Finite-size scaling plot of the data for $P(q)$ 
shown in Fig.~\ref{fig:P100}.
}
\label{fig:P100scal255}
\end{figure}

\begin{figure}[t]
\vspace*{6.8cm}
\includegraphics{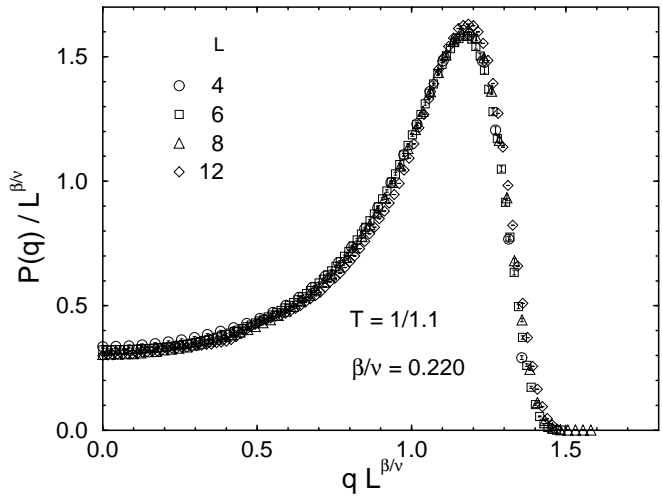}
  \caption{\protect\small
Finite-size scaling plot for $P(q)$ at $T \approx 0.8 T_c$.
}
\label{fig:P110scal220}
\end{figure}

Even though less detailed than the barrier data also the averaged
canonical probability densities $P(q) = [P_i(q)]_{\rm av}$ contain
important information on the system. Due to the average over all
realizations the $P(q)$ exhibit a double-peak shape; see
Figs.~\ref{fig:P088scal317}-\ref{fig:P110scal220}.
In principle the data stored in the time-series files 
allow canonical reweighting in a $\beta$-range around the simulation 
point. For disordered systems, however, where many realizations have to
be reweighted, great care is needed when estimating the valid
reweighting range. 

By analyzing the spin glass susceptibility, 
$\chi_{\rm SG} = N [\langle q^2 \rangle ]_{\rm av}$, we obtained in
Ref.~\cite{bj98} the best finite-size scaling fit
$\chi_{\rm SG} \propto L^{\gamma/\nu}$ at $\beta = 0.88$ with $\gamma/\nu
= 2.37(4)$ and a goodness-of-fit parameter $Q=0.25$. This was corroborated
by the curves of the Binder parameter, $g = (1/2) (3 - [\langle q^4
\rangle]_{\rm av}/[\langle q^2 \rangle ]_{\rm av}^2)$, which merge
around $\beta=0.89$. In the
low-temperature phase ($\beta > \beta_c$) the curves for different
lattice sizes seem to fall on top of each other, but our error bars
were still too large to draw a firm conclusion from this quantity.

Close to the transition temperature $T_c$ one expects that
the probability densities $P(q)$ for different lattice sizes
satisfy the finite-size scaling relation
\begin{equation}
P(q) = L^{\beta/\nu} \hat{P}(L^{\beta/\nu} q,L^{1/\nu}(T - T_c))\qquad,
\label{eq:P_scal}
\end{equation}
where $\hat{P}$ is a scaling function, and $\beta$ and $\nu$ are the
critical exponents of the order parameter and correlation length,
respectively. They are related to $\gamma/\nu = 2 - \eta$ by the standard
scaling relation $\beta/\nu = (d-\gamma/\nu)/2$. Using our estimate for
$\gamma/\nu \approx 2.37$ we thus obtain the value 
$\beta/\nu \approx 0.317$. 
Figure~\ref{fig:P088scal317} shows the probability densities $P(q)$
of the large-scale simulations, reweighted to the
transition temperature $T_c$ and rescaled according to 
(\ref{eq:P_scal}), using the above exponent estimates. We see that the data
for $L=4$, 6, and 8 fall almost perfectly onto a common master curve,
while the $L=12$ data obviously show some deviations, in particular
close to the peak. The overall appearance of Fig.~\ref{fig:P088scal317},
therefore, is actually worse than that of the corresponding plot in 
Ref.~\cite{bj98} which is based on a much smaller exploratory data set.
We strongly suspect that the reason for the deviations of the $L=12$ curve
can be traced back to problems with the admissible reweighting range for
our largest system size. Since for quenched, disordered systems 
the reweighting range for averaged quantities depends on the common overlap 
of the reweighting ranges for the individual realizations, it is indeed
conceivable that this problem does show up more dramatically for the larger
sample of realizations. Given the very good data collapse for the smaller 
lattices, and
with this explanation in mind, we feel that also our new
results are compatible
with the findings of Ref.~\cite{KaYo96}. We are
currently trying to improve the test of the finite-size scaling prediction
(\ref{eq:P_scal})
by redoing the simulations closer to $\beta_c$.
Narrowing the $q$-range may allow to simulate lattices of size $L=16$
and beyond.

The observation that the Binder parameter curves do not splay out at low
temperatures suggests that, similar to the 2d XY model, the correlation
length $\xi$ may be infinite -- or at least
larger than the simulated lattice sizes. Generalizing the second
argument of $\hat{P}$ to $L/\xi$ and assuming $L/\xi=0$, one thus
expects that the $P(q)$ should 
scale
also below $T_c$. With our simulation set-up, the most reliable test
of this conjecture can be
done at the simulation point $T=1 \approx 0.88 T_c$, since then
no temperature reweighting is involved. The raw data for $P(q)$ at
$T=1$ are shown in Fig.~\ref{fig:P100}. Notice
that $P(0)$ is slightly
growing with increasing system size.
By adjusting the
only free parameter, $\beta/\nu = 0.255$, we obtain the
finite-size scaling plot in Fig.~\ref{fig:P100scal255} which shows
a very clear data collapse onto a single master curve for all
lattice sizes.
Moreover, if we reweight our data to the even lower temperatures
$T=1/1.1$ (= $0.80 T_c$) and
$T=1/1.2$ (= $0.73 T_c$) we still find
a reasonable data collapse, see Fig.~\ref{fig:P110scal220}, but here
again extreme care is needed to
control the reliable reweighting range.

Of course, since our lattice sizes are relatively
small, we cannot conclude that the
correlation length is infinite below
$T_c$. If $\xi$ is large but finite, it is conceivable that 
we observe an
{\em effective} scaling behavior as long as $\xi > L$. We may conclude, that
the correlation length is unusually large ($\xi > 12$) down to $0.73 T_c$.
\section{Conclusions}
For the $3d$ EAI model at $\beta=1$ we have performed a high-statistics
calculation for probability distributions depending on the Parisi overlap
parameter~$q$. The results for free-energy barriers in $q$ became feasible
by using $q$-dependent (multi-overlap) weight factors in large-scale MC 
simulations. Although the tunneling performance is not optimal, the method
opens new horizons for spin glass simulations. 
Using slight modifications of the method (like narrowing the $q$-range,
including a magnetic field, etc.) will allow us to extend our investigation
into various interesting directions, like an improved study of the
thermodynamic limit at and below the freezing point, a study of the
4d EAI spin glass model, or
$\epsilon$-physics,
where one studies the influence of an interaction term~\cite{CaPa90}
$ \epsilon \sum_{i=1}^N s^1_i\, s^2_i = \epsilon\, N\, q $
in the Hamiltonian (\ref{energy}). In multi-overlap simulations we can 
obtain expectation values for arbitrary $\epsilon$-values by reweighting. 
Physically most interesting is the case where
a non-zero magnetic field is combined with a non-zero $\epsilon$-value.
%
    \vspace{0.6cm}\newline{\small 
W.J. thanks K. Binder for useful discussions and acknowledges support 
from the Deutsche Forschungsgemeinschaft
(DFG) through a Heisenberg Fellowship. The numerical simulations were
performed on T3E computers of CEA Grenoble, ZIB Berlin, and HLRZ J\"ulich.
We thank all institutions for their generous support. This
research was partially funded by the Department of Energy under Contracts
No. DE-FG02-97ER41022 and DE-FG05-85ER2500.
    }
    \end{document}